\relax
\catcode`\@=11
\def\slash{\@ifnextchar[{\@slash}{\@slash[\z@]}}
\def\@slash[#1]#2{\setbox\z@\hbox{$#2$}\@tempdima\wd\z@\box\z@%
\@tempdimb#1 \advance\@tempdimb-\@tempdima \kern\@tempdimb
\hbox to\@tempdima{\hss\@makeslash\hss}}
\def\@makeslash{$/$}			
\catcode`\@=12

\documentstyle[11pt]{article}

\relax
\catcode`\@=11

\newcommand{\ii}{{\rm i}}                               
\newcommand{\dd}{{\rm d}}                                
\newcommand{\vdd}[1]{{\frac{\delta }{\delta #1}}}        
\newcommand{\half}{{1 \over 2}}                           
\newcommand{\Lie}{{\cal L}}

\newcommand{\DD}{{\cal D}}
\newcommand{\doo}{{\partial}}

\newcommand{\Str}{{\rm Str \  }}

\newcommand{\ra}{\rightarrow}

\newcommand{\mibitem}[1]{\bibitem{#1}}
\newcommand{\bastar}{\begin{eqnarray*}}
\newcommand{\eastar}{\end{eqnarray*}}

\newcommand{\tensor}{\otimes}

\newcommand{\g}{{\bf g}}

\newcommand{\be}{\begin{equation}}
\newcommand{\ee}{\end{equation}}

\newcommand{\ba}{\begin{eqnarray}}
\newcommand{\ea}{\end{eqnarray}}

\begin{document}

\begin{titlepage}
\begin{flushright}
UU-ITP 27-1996 \\ hep-th/9612209
\end{flushright}

\vskip 0.5truecm

\begin{center}
{ \Large \bf 
Weil Algebras and Supersymmetry
\\

 }
\end{center}

\vskip 1.0cm

\begin{center}
{\bf Mauri Miettinen \\
\vskip 0.4cm
{\it Department of Theoretical Physics, Uppsala University
\\ P.O. Box 803, S-75108, Uppsala, Sweden \\}
\vskip 0.4cm}
\end{center}

\vskip 1.0cm

\rm
\noindent

We give a new interpretation for
the super loop space that has been used to formulate
supersymmetry. The fermionic coordinates in the super loop space
are identified as the odd generators of the Weil algebra. Their
bosonic superpartners are the auxiliary fields. The general
$N=1$ supermultiplet is interpreted in terms of Weil algebras.  As
specific examples we consider supersymmetric quantum mechanics,
Wess-Zumino model and supersymmetric Yang-Mills theory in four
dimensions.  Some comments on the formulation of constrained systems
and integrable models and non-Abelian localization are given.

\vfill

\begin{flushleft}
\rule{5.1 in}{.007 in}\\
$^{*}${\small E-mail: $~~$ {\small \bf
mauri@teorfys.uu.se, maumiett@rock.helsinki.fi} $~~~~$ \\
Research supported by Jenny and Antti Wihuri Foundation 
}
\end{flushleft}

\end{titlepage}

\section{Introduction}

Equivariant cohomology (for a review see eg. \cite{Moore}) arises when
a group acts on a manifold.  It yields a geometrical characterization
of the dynamics of various systems, eg. supersymmetric and topological
field theories.  Mathai-Quillen \cite{MQ} and Duistermaat-Heckman
\cite{DH} localization schemes also utilize equivariant cohomology.
In topological theories the BRST-operator has the structure of an
equivariant exterior derivative and the observables are equivariant
cohomology classes \cite{OSvB, equbrst}. In supersymmetric theories
the supersymmetry is generated by an equivariant derivative associated
to a circle action in a loop space \cite{palo}.  The
localization theorems, both finite dimensional and path integral
generalizations, also utilize the equivariant cohomology.

In this article we shall discuss different models for equivariant
cohomology and especially the the formulation of supersymmetric
theories. To do this we introduce the Weil algebra related to the
group acting on the manifold.  With the Weil algebra we can formulate
the Cartan, Weil and BRST-models \cite{kalkthesis}.  The fields of the
theories are divided to two classes. Half of the fields are
interpreted as the generators of the exterior algebra on the loop
space over function space. The other half generates the loop space
over the Weil algebra associated to the group action. This distinction
gives another geometric interpretation of supersymmetry different from
the super loop space construction \cite{palo}.

This article is organized as follows. We first represent models for
equivariant cohomology. The central concept is the Weil algebra which
is used to define the equivariant differentials. Then we show that by
adding a contractions along the circle action generators in the
relevant loop spaces we obtain supersymmetry operators. This
equivariant structure is shown to exist in a general $N=1$
supersymmetric theory by formulating the transformation rules of the
general $N=1$ multiplet \cite{Sohnius} with the equivariant
operators. As specific examples we discuss supersymmetric quantum
mechanics and four dimensional Wess-Zumino and super Yang-Mills
theories.

\section{Equivariant Cohomology and  Supersymmetry}

Let $G$ be a Lie group with Lie algebra $\g$ acting on a manifold
$M$. The action of $G$ is generated by vector fields $\{\chi_a, a
=1,... \dim G \}$ obeying the commutation relations of $\g$
\be\label{VectorLieAlgebra}
[ \chi_a, \chi_b ] = f_{ab}^c \chi_c \;.
\ee
Here we adopt the notation that group indices are from the beginning
and the space indices from the middle of the alphabet.  The
corresponding Lie derivatives $\Lie_a = [\dd , \iota_a ]_+$ obey the
same algebra. Here $\iota_a $ denotes the contraction along the
Hamiltonian vector field $\chi_a^k = \omega^{kl} \doo_l H_a$.  If the
$G$-action is symplectic so that it preserves the symplectic 2-form,
$\Lie_a \omega = \dd
\iota_a \omega =0$ and $H^1(M) = 0$ we can introduce a momentum map
$H(x): M \ra {\g}^*$ \cite{guiste} such that when one evaluates it on
a given Lie algebra element $\xi$ it yields the corresponding
Hamiltonian on $M$
\be\label{DefMomentumMap}
\langle  H(x) , \xi \rangle = H_{\xi} (x) \;.
\ee
This provides a one-to-one correspondence between vector fields
$\chi_a$ and corresponding components $H_a$ of the momentum map $H =
\phi^a H_a$ where $\{ \phi^a \}$ is the symmetric basis of
${\g}^*$. 
The functions $H_a$ satisfy the Poisson algebra
\be\label{Poisson}
\{H_a , H_b \} = \doo_k H_a \omega^{kl} \doo_l H_b  
  = f_{ab}^c H_c + \kappa_{ab}
\ee
where $\kappa_{ab} $ is a possible Lie algebra 2-cocyle. Equivariant
cohomology is roughly the de Rham cohomology of the quotient space
$M/G$. However, when $G$ does not act freely (it has fixed points),
$M/G$ is not a manifold. Therefore, a more precise definition is
required.

The topological definition uses universal bundles. The universal
bundle $EG \ra BG$ is such a bundle that every $G$-principal bundle
over $M$ is obtained by a pull-back: $P(G) = f^*(EG)$ for some $f: M
\ra BG.$ The universal bundle $EG$ is a contractible manifold with a 
free $G$-action and thus we can form the associated bundle $EG
\times_G M$. The topological equivariant cohomology is the de Rham
cohomology of $EG \times_G M /G$.

In this article we are more interested in the algebraic equivariant
cohomology since we are going to formulate supersymmetry using it.  To
do this we have to consider loop spaces which arise when we express
partition functions of supersymmetric theories (Witten indices) using
path integrals 
\be\label{FPI}
Z = \Str \exp[- \ii T {\cal H}] = \int_{PBC} \DD \Phi \exp( \ii S[\Phi] ) \;
\ee
where ${\cal H}$ is the Hamiltonian and $S[\Phi]$ the corresponding
action. All the fields $\Phi$ are assumed to have periodic boundary
conditions in time $\Phi(x, 0) = \Phi(x, T)$ and to vanish in
infinity. The fields are thus elements of the loop space $L\Phi$.

The classical model for $G$-equivariant cohomology is the Cartan model
in which we consider the equivariant exterior derivative for the
momentum map $H = \phi^a H_a$
\be\label{Cartan}
\dd_H =  \dd - \phi^a \iota_a
\ee
 which squares to $\dd_H^2 = - \phi^a \Lie_a$. Consequently, $\dd_H$
is nilpotent on the $G$-invariant subcomplex of differential forms
$\Lambda_G M$ and the equivariant cohomology is
\be\label{EqCartan}
H_G^* (M) = H^*( \Lambda_G M ) \;.
\ee

Previously it has been shown
\cite{palo} that $N=1$ supersymmetric theories can be interpreted in
terms of super loop space equivariant cohomology. This formulation
uses the Cartan model. In super loop space the coordinates $\varphi^k$
can be either Grassmann even or odd and the corresponding 1-forms
$\psi^k$ are thus odd and even, respectively.  By introducing
auxiliary fields a supersymmetric theory can be written in the form
(with implicit space-time integrations)
\be\label{SUSYCartan}
S[\Phi] = S_B + S_F = \int \vartheta_k \dot{\varphi}^k + \psi^k
\Omega_{kl} \psi^l = (\dd + \iota_{\dot{\varphi}} ) \vartheta =
\dd_{\dot{\varphi}} \vartheta \;.
\ee
Here $\vartheta$ is a pre-symplectic potential in the super loop space
and $\Omega$ the corresponding 2-form. The contraction
$\iota_{\dot{\varphi}}$ is along the vector field that generates the
circle action, or the translation in loop parameter $\tau$.  The
supersymmetry means that the action is equivariantly closed
\be\label{EqClosed}
 \dd_{\dot{\varphi}} (S_B + S_F) = \dd_{\dot{\varphi}}^2 \vartheta =
\Lie_{\dot{\varphi}} \vartheta \sim 0 \;.
\ee
The Weyl identity
\be\label{Translation}
\dd_{\dot{\varphi}}^2 = \Lie_{\dot{\varphi}} = \dd /\dd \tau
\ee
 provides a representation of the supersymmetry algebra.  The super
loop space structure has been shown to be general by considering the
transformation laws of a general $N=1$ supermultiplet \cite{palo}.

To define the Weil and BRST-models for equivariant cohomology we
introduce the Weil algebra $W(\g) = S({\g}^*) \tensor
\Lambda({\g}^*)$ where $S({\g}^*)$ is the symmetric algebra on ${\g}^*$ and
$\Lambda({\g}^*)$ the exterior algebra on ${\g}^*$. The symmetric
algebra $S({\g}^*)$ is generated by commuting basis elements $\{\phi^a
\}$ and $\Lambda({\g}^*)$ by anticommuting elements $\{\eta^a \}$. The
Weil differential $\dd_W$ acting on $W(\g) $ is given by
\ba\label{WeilDiff}
 \dd_W \eta^a &=& \phi^a - \half f_{bc}^a \eta^b \eta^c  \;, \cr
 \dd_W \phi^a &=& - f_{bc}^a \eta^b \phi^c \;.
\ea
The Weil differential is nilpotent with a trivial cohomology.  The
complex of the Weil and BRST-models is the tensor product $\Lambda M
\tensor W(\g)$.

In the Weil model we introduce the nilpotent differential on $\Lambda
M \tensor W(\g)$
\be\label{TotalDiff}
\dd_T = \dd  +  \dd_W \;.
\ee
The equivariant cohomology is given by restricting to basic forms
which are $G$-invariant and horizontal. To do this we introduce an
interior product and a Lie derivative on $W(\g)$ by
\ba\label{WeilLie}
I_a \eta^b & =& \delta_a^b \;, \cr
I_a \phi^b & = & 0 \;, \cr
L_a & =& [\dd_W, I_a]_+ \;.
\ea
Forms annihilated by ($\Lie_a +L_a$) ($G$-invariant) and ($\iota_a
+I_a$) (horizontal) are elements of the Weil model for $G$-equivariant
cohomology.

The BRST-model differential is obtained by defining the nilpotent
differential on $ W(\g) \tensor \Lambda M$
\be\label{BRST-operator}
s = \dd  +  \dd_W + \eta^a  \Lie_a - \phi^a \iota_a \;.
\ee
 The BRST- and Weil model differentials are
related by a conjugation
\cite{kalkthesis}
\be\label{Conjugation} 
s = \exp( - \eta^a \iota_a ) \dd_T \exp( \eta^b \iota_b) \;.
\ee 
Since the cohomology of $\dd_W$ is trivial the cohomology of $s$ is
the deRham cohomology of $M$. The BRST-model for equivariant
cohomology is given by the cohomology of $s$ when it is restricted to
basic forms. The reason for calling $s$ to be a BRST-differential is
that eg. in topological Yang-Mills the structure of the BRST operator
is exactly the same \cite{OSvB, equbrst}. The fields $\eta^a$ are the
ghosts for the constraints $\Lie_a$ and $\phi^a$ are the ghosts for
ghosts required for first stage reducible theories.

Now we are going to show that a similar interpretation of
supersymmetry exists also in the context of the Weil and
BRST-models. In fact, the role of the odd coordinates in the super
loop space becomes more precise: they are the odd generators $\eta^a$
of the loop space of the Weil algebra $LW(\g).$ Also the auxiliary
fields that do not have any dynamics but are introduced to have a
balance between the number of even and odd fields are the generators
$\phi^a$ of $LW(\g)$. To get dynamics to the fields we add
contractions along circle action on $L\Phi$ and $LW(\g)$. The
integration over space-time is implicit in the following
\ba\label{Extensions}
\dd_T &\ra& Q_\tau  =  \dd + \dd_W  +\iota_{\dot{\varphi}} +
I_{\dot{\eta}} \;, \cr 
s &\ra& s_\tau = \dd + \dd_W - \phi^a \iota_a +\eta^a \Lie_a
+\iota_{\dot{\varphi}} + I_{\dot{\eta}} \;.
\ea
These generators square to 
\be\label{SUSYRep}
Q^2_\tau = s_\tau^2 = {\dd \over \dd \tau} \;.
\ee
Here $\tau$ is the circle parameter which is not necessarily time but
can be a light-cone coordinate as well. The relation (\ref{SUSYRep})
also provides a representation of the supersymmetry algebra.  The
operators $Q_\tau$ and $s_\tau$ are thus supersymmetry generators.
The Weil model generator has the explicit (Lagrangian) realization
(now $\varphi^k$ is even and $\psi^k \sim \dd \varphi^k$ odd)
\ba
Q_{\tau} & = & \psi^k {\delta \over \delta \varphi^k} + (\phi^a -
\half f^a_{bc} \eta^b \eta^c ) {\delta \over \delta \eta^a} +
\dot{\varphi}^a   {\delta \over \delta
\psi^k} + \dot{\eta}^k  {\delta \over \delta \phi^k } \;. 
\ea
A Hamiltonian realization for the BRST-generator is obtained by
introducing the conjugate variables with non-trivial Poisson brackets
\ba
\{ p_l , \varphi^k  \} & = &\{ \psi^k , \bar{\psi_l}  \} = \delta^k_l \;, \cr
\{ \eta^a , {\cal P}_b   \} & = & \{ \phi^a, \pi_b     \} = \delta^a_b \;.
\ea
The BRST-differential is then
\ba
s_{\tau} & = & \psi^k p_k + ( \phi^a -
\half f^a_{bc} \eta^b \eta^c ) {\cal P}_a - \phi^a \iota_a + \eta^a \Lie_a +
\dot{\varphi}^k \bar{\psi}_k + \dot{\eta}^a  \pi_a \;.
\ea

Supersymmetric actions can be obtained by acting with $Q_\tau$ and
$s_\tau$ on a symplectic potential 
$
\Theta$ on  $LW(\g) \otimes L\Phi$
since every action of the form $S = Q_\tau \Theta$ or $S = s_\tau
\Theta$ is automatically supersymmetric in view of (\ref{SUSYRep}). 
However, renormalizability restricts possible symplectic
potentials.  In the theories that we are going to consider the
potential is
\be\label{ThePotential}
\Theta= \vartheta_k \psi^k + \half \phi^a \eta^a - \eta^a W_a
\ee
 where $W_a$ is the superpotential of the model, as we shall see in
specific examples.  In the Weil model we obtain the action
\ba\label{WeilAction}
 S_W &=& Q_\tau \Theta = \int \vartheta_k \dot{\varphi}^k +  \half \eta^a
\dot{\eta}^a + \half \Omega_{kl} \psi^k \psi^l \cr 
&+&\half \phi^a \phi^a -  \half f^a_{bc} \phi^a \eta^b \eta^c 
 - \phi^a W_a - \eta^a \doo_k W_a \psi^k - \half f^a_{bc} \eta^b
\eta^c W_a \;.
\ea
In the BRST-model a supersymmetric action becomes
\ba\label{BRST-action}
S_{B} &=& s_\tau \Theta =
\int \vartheta_k \dot{\varphi}^k + \eta^a \dot{\eta}^a - \phi^a
H_a +\half \Omega_{kl} \psi^k \psi^l \cr 
 &+& \half \phi^a (\phi^a - {3 \over
2} f^a_{bc}  \eta^b \eta^c ) - \phi^a W_a - \eta^a \doo_k W_a
\psi^k - \eta^a \{H_a, W_b \} 
\eta^b \;.
\ea
Here we assume that $\Lie_a \vartheta = 0$ which allows us to identify
$H_a = \iota_a \vartheta.$ The Weil algebra generator $\phi^a$ appears
purely algebraically in the actions and is therefore an auxiliary
field. Also, the interaction that involves the superpotential produces
a coupling between $L\Phi$ and $LW(\g)$.  In the Weil model we have
only trivial circle action in the loop spaces. By choosing suitable
Hamiltonians $H_a$ in the BRST-model we expect to get some non-Abelian
generalizations for supersymmetric models with a non-trivial
Hamiltonian flow in the loop space. These models might be relevant for
non-Abelian localization
\cite{wtwodim}.  Another interpretation \cite{equbrst} 
for (\ref{BRST-action}) is that it is a constrained system with a
constraint algebra (\ref{Poisson}) which is encompassed by the
BRST-operator.

In the supersymmetric models that we consider it is sufficient to use
the Abelian Weil model differential with $f^a_{bc} = 0 $. The Weil
model action reduces to
\ba
S_W & = & \int \vartheta_k \dot{\varphi}^k + \half
\eta^a \dot{\eta}^a + \half \Omega_{kl} \psi^k \psi^l \cr 
&+&\half \phi^a \phi^a 
 - \phi^a W_a - \eta^a \doo_k W_a \psi^k \;.
\ea
An interesting prospect is to consider the BRST-model with an
Abelian Weil algebra. In this case the Hamiltonians are in involution
and the action
\ba
S_B & = & \int \vartheta_k
\dot{\varphi}^k + \half \eta^a \dot{\eta}^a - \phi^a H_a +\half
\Omega_{kl} \psi^k \psi^l \cr
&+& \half \phi^a \phi^a -\phi^a W_a - \eta^a
\doo_k W_a \psi^k - \eta^a \{H_a, W_b \}\eta^b \;.
\ea
might represent some integrable models.

\section{The  General N=1 Supermultiplet in Four Dimensions}

To show that the Weil algebra structure is present in general $N=1$
supersymmetric theories in four dimensions we formulate the
transformation laws of the general $N=1$ supermultiplet \cite{Sohnius}
using the Weil model supersymmetry generator.  The multiplet consists
of the following fields: scalar $M$, pseudoscalars $C,N,D$, vector $A_{\mu}$
and Dirac spinors $\chi,
\lambda$.  The transformation properties of the multiplet with 
a Grassmann spinor parameter $\zeta$ are
\ba
\delta C & = & \bar{\zeta} \gamma_5 \chi  \;, \cr
\delta \chi & = & ( M+ \gamma_5 N) \zeta - \ii \gamma^{\mu} (A_{\mu} +
\gamma_5 \doo_{\mu} C) \zeta \;, \cr
\delta M & =  & \bar{\zeta} ( \lambda - \ii \slash{\doo} \chi ) \;, \cr
\delta N & = & \bar{\zeta} \gamma_5 ( \lambda -\ii \gamma_5
\slash{\doo}  \chi ) \;, \cr
\delta A_{\mu} & = & \ii \bar{\zeta}  \gamma_{\mu} \lambda +
\bar{\zeta} \doo_{\mu} \chi \;, \cr
\delta \lambda & = & - \ii \sigma^{\mu \nu} \zeta \doo_{\mu} A_{\nu} -
\gamma_5 \zeta D \;, \cr
\delta D & = & - \ii \bar{\zeta} \slash{\doo} \gamma_5 \lambda \;.
\ea
These transformations are generated by the Majorana supercharge
\be
Q = \pmatrix{ &Q_\alpha & \cr
              &\bar{Q}^{\dot{\alpha}} &     }
\ee
which obeys the anticommutator
\be
[Q , Q ]_+ = 2 (\gamma^\mu C ) P_\mu \;.
\ee
Using the Majorana representation $\gamma^0 = - \sigma^2 \otimes 1, ~
\gamma^1 = - \ii \sigma^3 \otimes \sigma^1, ~ \gamma^2 = \ii
\sigma^2 \otimes 1, ~ \gamma^3 = - \ii \sigma^3 \otimes \sigma^3$
we have the following relevant entries
\be
  (\gamma^\mu C ) P_\mu = \pmatrix{ & \ii \doo_+ & * & * & * & \cr 
                                   & *  & \ii \doo_+ & * & * & \cr 
                                     & * & *  & \ii \doo_- & * & \cr
                                    & * & * & *  & \ii \doo_- & } \;.
\ee
Here $\doo_\pm$ denote derivatives with respect to light cone 
coordinates $x^\pm = x^2 \pm t$.

For simplicity we choose to consider the supersymmetry transformations
generated by the component $Q_1$.  The following redefinitions of the
fields simplify the transformation rules (for details see \cite{palo})
\begin{eqnarray}
M' & = & M+A_{z}+\partial_{x}C\, , \nonumber\\
N' & = & N+A_{x}-\partial_{z}C\, , \nonumber\\
\lambda_{1}' & = & \lambda_{1}-\partial_{z}\chi_{1}\, , \nonumber\\
\lambda_{2}' & = & \lambda_{2}-
\partial_{x}\chi_{1}\, , \label{redef}\\
\lambda_{3}' & = & 2\lambda_{3}-\partial_{-}\chi_{1}\, , \nonumber\\
D' & = & D+\partial_{x}A_{z}-\partial_{z}A_{x} \;.  \nonumber
\end{eqnarray}
The transformation rules for the primed fields are
\begin{eqnarray}
Q_{1} C & = & \chi_{2}\ ,\nonumber\\
Q_{1} (\chi_{1}, \chi_{2}, \chi_{3}, \chi_{4})\;, 
 & = &  (\ii A_{+}, \ii \partial_{+}C, \ii M', \ii N')\; ,\nonumber\\
Q_{1} M' & = & \partial_{+}\chi_{3}\; , \nonumber\\
Q_{1} N' & = & \partial_{+}\chi_{4}\; , \label{trans2}\\
Q_{1} (A_{+}, A_{-}, A_{x},
A_{z}) & = &  (\partial_{+}\chi_{1}, -\lambda_{3}, -\lambda_{2},
-\lambda_{1})\; ,\nonumber\\
Q_{1} (\lambda_{1}', \lambda_{2}', \lambda_{3}', \lambda_{4}) & = &
(-\ii \partial_{+}A_{z}, -\ii \partial_{+}A_{x}, -\ii
\partial_{+}A_{-}, -\ii D') \; ,\nonumber\\
Q_{1} D' & = & -\partial_{+}\lambda_{4}\; .\nonumber
\end{eqnarray}

The exterior derivative and the Weil differential, as well as the
contractions can now be read from the transformation laws. The bosonic
fields $\varphi^k \sim (C, A_-, A_x, A_z )$ can be viewed as the
coordinates in $L\Phi$ while the fermionic fields
$\psi^k \sim (\chi_2, -\lambda_3', - \lambda_2', -\lambda_1')$ are
their differentials. We thus identify
\be
{\rm d}  =  \chi_{2}\vdd{C} -\lambda_{3}'\vdd{A_{-}}-
\lambda_{2}'\vdd{A_{x}}-\lambda_{1}'\vdd{A_{z}}
\ee
as the exterior derivative in $L\Phi$. The Weil algebra $LW(\g)$ is
generated by the fermionic fields $\eta^a \sim (\chi_1 , \chi_3, \chi_4,
\lambda_4' )$. According to the rules they transform to their 
 bosonic  superpartners
$\phi^a \sim (\ii A_+ ,\ii M', \ii N', -\ii D')$. This is exactly the
transfromation generated by the Weil differential $\dd_W$ which
becomes
\be
\dd_W = \ii A_+ \vdd{\chi_1}  + \ii M'\vdd{\chi_{3}}+
\ii N'\vdd{\chi_{4}} -  \ii D'\vdd{\lambda_{4}}   \;.
\ee
The remaining transformation laws of the multiplet are such that the
fields transfrom to space-time derivatives. These rules are generated
by the contractions on $L\Phi$ and $LW(\g)$
\ba
\iota_+  & = &  \ii \partial_+C \vdd{\chi_{2}} -\ii \partial_{+}A_- 
\vdd{\lambda_{3}'}
-\ii \partial_{+}A_{x} \vdd{\lambda_{2}'}
-\ii \partial_{+}A_{z}\vdd{\lambda_{1}'} \cr I_+ & =
&\partial_{+}\chi_{3}\vdd{M'} +\partial_{+}\chi_{4}\vdd{N'} -
 \partial_{+}\lambda_{4} \vdd{D'} + \doo_+ \chi_1 \vdd{A_+}  \, .
\ea
The Weil model supersymmetry generator is the sum of these terms
\be
Q_+ = \dd + \dd_W + \iota_+ + I_+ \;.
\ee
It represents the superalgebra since  $Q_+^2 = \ii \doo_+$.

\section{Specific Theories}

We shall now show by some examples that $N=1$ supersymmetric theories
have the Weil algebra structure by explicitly representing the Weil
generator and the symplectic potential. We first concentrate on chiral
fields that describe matter fields by considering the supersymmetric
quantum mechanics and the four dimensional Wess-Zumino-model. Then we
discuss the four dimensional supersymmetric Yang-Mills theory without
matter as an example of a model which involves vector superfields.

The supersymmetric quantum mechanics has the action
\ba\label{SQM}
S &=& \int \half \dot{q}^2 + \half (\theta_1 \dot{\theta}_1 + \theta_2
\dot{\theta}_2) - \half W_q^2 - \theta_2 W_{qq} \theta_1 \cr
& =& \int  \half \dot{q}^2 + \half \theta_1 \dot{\theta}_1 +\half  \theta_2
\dot{\theta}_2) + \half F^2   -  \half F W_q - \theta_2 W_{qq} \theta_1
\ea
where $W_q = \doo W / \doo q $ is the superpotential and $F$ an
auxiliary field. We notice that the action is of the form
(\ref{WeilAction}) with the following identifications. The terms
$\half \dot{q}^2 \sim \vartheta_k \dot{\varphi}^k$ and $\half \theta_1
\dot{\theta}_1 \sim \half \eta^a \dot{\eta}^a$ are the 
pre-symplectic potentials on $L\Phi$ and $LW(\g)$ respectively. The
corresponding 2-forms are $\half \theta_2 \dot{\theta}_2 \sim
\half \Omega_{kl} \psi^k \psi^l$ and $\half F^2$. The interaction part
 $ S_{int} = -F W_q - \theta_2 W_{qq} \theta_1 \sim -\phi^a W_a -
\eta^a \doo_k W_a \psi^k$ appears as a coupling between the
 loop spaces $L\Phi$ and $LW(\g)$.
The exterior derivative on $L\Phi$ is $ \dd = \theta_1 \delta / \delta
q$ and the Weil differential $\dd_W = F \delta / \delta
\theta_2$. The contractions are $\iota_{\dot{q}} = \dot{q} {\delta / \delta
 \theta_1}$ and $I_{\dot{\theta}_2} = \dot{\theta}_2 {\delta / \delta
 F}$. The Weil model supersymmetry generator is given by
\be
Q_t = \theta_1 \vdd{q} + F \vdd{\theta_2} + \dot{q} \vdd{\theta_1} +
\dot{\theta}_2 \vdd{F}
\ee
and the action is obtained from the symplectic potential
\be
\Theta = \half \dot{\theta}_1 q + \half F \theta_2 - \theta_2 W_q
\ee
which coincide with the form in the general discussion.

This generalizes to the four dimensional Wess-Zumino model with the
action
\ba\label{4DWZ}
S &=& \int \doo_{\mu} A^* \doo^{\mu} A - {1 \over 4} W_{A} W_{A^*} +
{\ii \over 2 } \bar{\psi} \gamma^{\mu } \doo_{\mu} \psi \cr
& +& {1 \over 4} \bar {\psi} (W_{A A} - W_{A^* A^*} ) \psi + {1 \over 4}
\bar{\psi} \gamma^5 (W_{A A} + W_{A^* A^*} ) \psi 
\ea
The Majorana representation of the Dirac matrices is $\gamma^0 =
\sigma^1 \otimes 1, \gamma^k = - \ii \sigma^2 \otimes \sigma^k,
\gamma^5 = \sigma^3
\otimes 1$. The charge conjugation is $C = - \ii \sigma^3 \otimes
\sigma^2$ and the Majorana spinor becomes
\be\label{$DMajorana}
\psi = \pmatrix{ &\theta_1 & \cr &\theta_2 & \cr & \bar{\theta}_2 &
\cr & - \bar{\theta}_1 & }
\ee

To write the Wess-Zumino action in the desired form we introduce
light-cone coordinates $x^{\pm} = x_0 \pm x_3$ and a complex
coordinate $z = x +\ii y$ and bosonic auxiliary fields $F, F^*$. Using
these the action can be written as
\ba\label{4DAux}
S &= &\int 2 \doo_- A^* \doo_+ A - \half F F^* - {\ii \over 2}
F ( W_{A} +2 \bar{\doo} A ) - {\ii \over 2}
F^* (W_{A} + 2 \doo A^* ) \cr
&+& \ii \bar{\theta_1} \doo_-
\theta_1 +  \ii \bar{\theta_2} \doo_- \theta_2 - \ii \bar{\theta_2} \doo_+
\theta_2 - \ii \bar{\theta_2} \bar{\doo} \theta_1 - \ii \bar{\theta}_1 \doo
\theta_2  \cr
&+&  \half W_{A A} \theta_1 \theta_2 - 
\half  W_{AA}^* \bar{\theta}_1 \bar{\theta}_2
\ea
We choose $(A, A^*) \sim \varphi^k$ as coordinates and $(\theta_1,
\bar{\theta}_1) \sim \psi^k $ as corresponding 1-forms on $L\Phi$. The Weil
algebra is generated by coordinates $\eta^a \sim (\theta_2,
\bar{\theta}_2)$ and 1-forms $\phi^a \sim(F,F^*)$. We choose the 
circle action in the light-cone direction $x^+$. The Weil model 
supersymmetry generator is then
\ba\label{4DGen}
Q_+ &=& \ii \theta_1 {\delta \over \delta A} + \ii \bar{\theta}_1
{\delta
\over \delta A^*} + \pi {\delta \over \delta \theta_1} + \pi^* {\delta
\over \delta \bar{\theta}_1 } \cr
&+& \doo_+ A {\delta \over \delta
\theta_2 } + \doo_+ A^*  {\delta \over \delta \bar{\theta}_2} +
\ii \doo_+ \theta_2 {\delta \over \delta \pi} + \ii \doo_+\bar{\theta}_2
{\delta \over \delta \pi^*}
\ea
which squares to $Q_+^2 = \ii \doo_+$. The action is obtained from the
symplectic potential
\be\label{4DPotential}
\Theta = \half (\doo_-A \bar{\theta}_1 + \doo_- A^* \theta_1 ) - \half (
\theta_2 \pi + \bar{\theta}_2 \pi^* ) + {\ii \over 2} (\doo A^* + W_A )
\theta_2 + {\ii \over 2} ( \bar{\doo} A + W_{A}^* ) \bar{\theta}_2 \;.
\ee
The interpretation of various terms is as in the supersymmetric
quantum mechanics.

Finally, the action of the supersymmetric Yang-Mills theory is
\be\label{SYMAction}
S =  \int - {1 \over 4} F^a_{\mu \nu} F^{a \mu \nu} + {\ii \over 2}
\bar{\psi} \gamma^{\mu} D_{\mu} \psi 
\ee
where $F^a_{\mu \nu} = \doo_{\mu} A^a_{\nu} - \doo_{\nu} A^a_{\mu}
+f^{abc} A^b_{\mu} A^c_{\nu }$ is the field strength of the gauge
field $A^a_{\mu}$, $\psi^a $ is a Majorana spinor in the adjoint
representation of the gauge group and $D_{\mu}^{ab} = \delta^{ab}
\doo_{\mu} + f^{abc} A^c_{\mu} $ is the covariant derivative. To write
 this action in the required form we first
consider the Hamiltonian form of the action
\be\label{SYMHam}
S = \int E^a_i \dot{A}^a_i - \half E^a_i E^a_i - \half B^a_i B^a_i +
{\ii \over 2} \bar{\rho}^a \dot{\rho}^a + {\ii \over 2} \bar{\eta}^a
\dot{\eta}^a + \rho^a \sigma^i  D^{ab}_i \eta^b
\ee
which is subject to the Gauss law ${\cal G}^a = D^{ab}_i E^b_i + { \ii
\over 2} f^{abc} (\rho^{\dagger b} \rho^c + \eta^{\dagger b} \eta^c )
\sim 0$. Here we have defined the spinors
\ba
\rho^a &= & \pmatrix{& \theta^a_1 & \cr 
	         & \bar{\theta}^a_1 & } \;,  ~~~~~~~
\eta^a =  \pmatrix{& \theta^a_2 & \cr
	         & \bar{\theta}^a_2 & } \;.
\ea
The change the variables $E^a_i \ra E^a_i + \dot{A}^a_i$ transforms the action to a form which admits an interpretation in terms of the Weil model:
\be
S = \int \half (\dot{A}^a_i )^2 - \half (E^a_i)^2 + \ii E^a_i B^a_i +
{\ii \over 2} \rho^{\dagger a} \dot{\rho}^a + {\ii \over 2}
\eta^{\dagger a}
\dot{\eta}^a + \rho^a \sigma^i  D^{ab}_i \eta^b \;.
\ee
The term $\half (\dot{A}^a_i)^2 \sim \vartheta_k \dot{\varphi}^k$ is
the pre-symplectic potential. The electric field $E^a_i \sim \phi^a$
is an auxiliary field. The second term is therefore the
symplectic 2-from on $LW(\g)$. The symplectic 1-forms on $L\Phi$ and
$LW(\g)$ are thus $\rho^a \dot{\rho}^a$ and $\eta^a \dot{\eta}^a $,
respectively. The magnetic field $B^a_i = \half \epsilon_{ijk} F^{a
jk}$ appears as the superpotential in this approach.

According to our standard recipe we can write the Weil model
supersymmetry generator, which in this case is spinorial. The circle
action is parameterized by the time:
\be
Q_t^{\beta} = \rho^a_{\alpha} \sigma^i_{\alpha \beta} {\delta \over
\delta A^a_i } + E^a_i \sigma^i_{\alpha \beta} {\delta \over \delta
\eta^a_{\alpha}} + {3 \ii \over 2} \dot{A}^a_i \sigma^i_{\alpha \beta}
{\delta \over \delta \bar{\rho}^a_{\alpha}} + {2 \ii \over 3}
\dot{\bar{\eta}}^a_{\alpha} \sigma^i_{\alpha \beta} {\delta \over
\delta E^a_i } \;.
\ee
Here $\sigma^i$ are the Pauli matrices.  The action is then obtained
from the symplectic potential of the form (\ref{ThePotential})
\be
\Psi_{\delta} = {\ii \over 6} \bar{\rho}^c_{\gamma} \sigma^k_{\gamma \delta}
\dot{A}^c_k - {1 \over 4} \eta^c_{\gamma} \sigma^k_{\gamma \delta}
E^c_k + {\ii \over 2} \eta^c_{\gamma} \sigma^k_{\gamma \delta} B^c_k \;.
\ee

\section{Conclusions}

In conclusion, we have applied Weil model for equivariant cohomology
to give a geometrical interpretation for $N= 1$ supersymmetry. In the
Lagrangian formulation we have shown that supersymmetry can be
interpreted in terms of bosonic loop space equivariant cohomology and
Abelian Weil algebras that are introduced in the Weil model.  In this
formulation half of the fermionic fields were interpreted as
differentials $\psi^k$ for half of the bosonic fields $\varphi^k$. The
other half of the fermions were interpreted as coordinates $\eta^a$ on
the Weil algebra $LW(\g)$. Their superpartners, the auxiliary fields,
were were the one forms $\phi^a$ on $LW(\g)$.  This identification
made the role of anticommuting coordinates and commuting 1-forms in
the super loop space quite clear: they are the odd and even generators
of the Weil algebra, respectively. The dynamics of the fields was
given by contractions along circle actions in the loop spaces.

We also discussed some aspects of the BRST-model for equivariant
cohomology. The BRST-differential has the structure of a BRST-operator
in constrained systems. We thus propose that many supersymmetric
constained systems can be realized by our formalism. Non-Abelian Weil
algebras with Hamiltonians $H_a$ could produce non-trivial Hamiltonian
flow in the loop space. This might give some interesting
generalizations for supersymmetric theories. With Hamiltonians $H_a$
in involution the systems might be integrable.

The Weil algebra structure was shown to be general general in
supersymmetry by considering the transformation rules of the
general $N= 1 $ supermultiplet. As examples we analyzed
supersymmetric quantum mechanics, and four dimensional Wess-Zumino and
super Yang-Mills theories with the Weil model.

 Some prospects are in developing our construction to extended
supersymmetric theories and integrable models. It would also be
interesting to consider theories in which we have non-Abelian Weil
algebra structures instead of Abelian which we have discussed. This
might be relevant for non-Abelian localization of path integrals.

\vspace{1.0 cm}
{\bf Acknowledgments}

M.M. thanks A.J. Niemi for introducing this problem and for comments
on the manuscript as well as Jenny and Antti Wihuri Foundation for
support.

 \end{document}